\documentclass[structabstract]{aa}

\usepackage{graphicx, rotate, lscape}
\usepackage{natbib}
\usepackage{latexsym,amssymb,verbatim}

\begin{document}
  \title{Molecular Tracers of Filamentary CO Emission Regions
  Surrounding the Central Galaxies of Clusters}

  \author{E. Bayet\inst{1}
    \and
    T. W. Hartquist\inst{2}
    \and
    S. Viti\inst{1}
    \and
    D.A. Williams\inst{1}
    \and
    T. A. Bell\inst{3}
  }

  \institute{Department of Physics and Astronomy,
    University College London, Gower Street,
    London WC1E 6BT, UK.\\
    \email{eb@star.ucl.ac.uk, daw@star.ucl.ac.uk}
    \and
    School of Physics and Astronomy
    University of Leeds
    Leeds LS2 9JT\\
    \email{twh@ast.leeds.ac.uk}
    \and
    Department of Astronomy,
    California Institute of Technology,
    Pasadena, CA 91125, USA\\
    \email{tab@phobos.caltech.edu}
  }

   \date{Received June 16, 2009; accepted June 16, 2009}

  \abstract
   {Optical emission is detected from filaments around the central galaxies of clusters of galaxies. These
filaments have lengths of tens of kiloparsecs. The emission is possibly due to heating caused by the
dissipation of mechanical energy and by cosmic ray induced ionisation. CO
  millimeter and submillimeter line emissions as well as H$_{2}$ infrared
  emission originating in such filaments surrounding NGC~1275, the central galaxy of the Perseus
  cluster, have been detected.}
   {Our aim is to identify those molecular species, other than CO,
  that may emit detectable millimeter and submillimeter line features
  arising in these filaments, and to determine
  which of those species will produce emissions that might serve as
diagnostics of the dissipation and cosmic ray induced ionisation.}
   {The time-dependent UCL photon-dominated region modelling code was
  used in the construction of steady-state models of molecular
  filamentary emission regions at appropriate pressures, for a range of
  dissipation and cosmic ray induced ionisation rates and incident
  radiation fields.}
   {HCO$^+$ and C$_2$H emissions will potentially provide information
  about the cosmic ray induced ionisation rates in the filaments. HCN
  and, in particular, CN are species with millimeter and submillimeter
  lines that remain abundant in the warmest regions containing molecules.}
   {Detections of the galaxy cluster filaments in HCO$^{+}$, C$_{2}$H, and CN
  emissions and further detections of them in HCN emissions would
  provide significant constraints on the dissipation and cosmic ray
  induced ionisation rates.}


   \maketitle
%

\section{Introduction}

The central galaxies of many clusters of galaxies are surrounded
by filaments observed in optical emission lines to extend up to
many tens of kiloparsecs (e. g. \citealt{Craw99, Cons01}).
The filaments in NGC~1275, the central galaxy of the Perseus
cluster, are typically of the order of 100 pc thick.
\citet{McNa96} suggested that the filaments originate in
interactions between the hot intracluster medium and the
relativistic plasma associated with the AGN-driven radio features
found in and around the central galaxies. \citet{Pope08b} modelled
the filaments of NGC~1275 as material ablated from clouds and dragged by a more
diffuse outflow moving away from NGC~1275 at more than $10^3$ km
s$^{-1}$. \citet{Pope08a} suggested that dissipation occurring
during the momentum transfer from the diffuse flow to the
filaments might power the filamentary emission, an idea related to
that of \citet{Craw92} that turbulent mixing layers play a role in
generating the emission. \citet{Ferl08, Ferl09} have constructed
detailed models of the optical and infrared line emission of
filamentary gas heated by similar dissipation and by cosmic ray
induced ionisation. They were motivated in part by the
difficulties encountered in attempts to match the observed spectra
with the results of models in which photoionisation by stellar
radiation is assumed to be the dominant heating mechanism (e. g.
\citealt{John07}).

Infrared H$_{2}$ emission lines trace the optical emission of the
filaments \citep{Jaff05}. \citet{Salo06, Salo08b} obtained CO(1-0),
and CO(2-1) single dish data on NGC~1275 and its surrounding filaments and they observed HCN(3-2) at the position of 3C84, the radio source located at the centre of the galaxy NGC~1275. These two sets of observations established that both regions contain cold molecular gas. \citet{Lim08, Salo08a,
  Ho09} have obtained interferometric observations of the CO(2-1)
emission of some of the NGC~1275 filaments.

The required heating rates per particle due to dissipation and/or
cosmic ray induced ionisation found by \citet{Ferl09} are several
orders of magnitude higher than those in the nearest Milky Way Giant
Molecular Clouds, and the required thermal pressures identified by
\citet{Ferl09} (see also \citealt{Sand04, Sand07}) are about an order of
magnitude higher than those of
the envelopes of dense cores associated with low mass star
formation. Though the heating and ionisation rates in the cold
molecular regions of the NGC~1275 filaments may be lower than those in
the corresponding optical emission regions, they still may greatly
exceed those relevant to the nearby Milky Way dense cores. If these rates
are
relatively high, the chemical composition in the NGC~1275 filaments
should differ from that in a typical nearby dense core and may provide
a diagnostic of the conditions in the filaments.

Hence, we have examined
the chemistry of molecular regions subject to heating rates per volume
and cosmic ray ionisation rates per particle in ranges that extend up
to those approaching those that, according to \citet{Ferl09}, obtain
in the optical emission line regions of the NGC~1275 filaments. Our
purpose is to identify those molecular species that may be detectable and
to determine which of those species may act as diagnostics of the
dissipation heating and of the cosmic ray intensity in the filaments. We
shall consider a range of parameters up to those indicated by the
observations; however, it is not our purpose here to model the filaments in
NGC~1275.

Section \ref{sec:model} contains details of the physical and
chemical models that we adopted to describe the molecular component of
the filaments. In section \ref{sec:resu} we present our results, and
section \ref{sec:conclu} ends the paper with a discussion and
some conclusions.


\section{The Models}\label{sec:model}

We compute the filament chemistry by assuming that the filament may be
represented by a steady-state photon-dominated region (a PDR) with
constant thermal pressure. To solve for the chemistry in the filament we
employ the UCL\_PDR time-dependent code, as implemented by \citet{Bell05, Bell07} and \citet{Baye09a} run for a sufficiently long time to reach steady state. This code operates in one space
dimension and computes self-consistently the chemistry and the temperature
as functions of depth and time within a semi-infinite slab, taking
account of a wide range of heating and cooling processes. In the present
work, the code is used to determine the chemical and thermal properties at
all depths up to a maximum of about 20 visual magnitudes. For the present work, we have updated the UCL\_PDR code to include additional radiative
cooling due to rotational transitions of $^{13}$CO, C$^{18}$O, CS, OH and
H$_2$O. The escape probability formalism of \citet{DeJo80} is used to
determine non-LTE level populations and resulting line intensities at each
depth- and time-step, in the same manner as for the existing coolants in the
code. Collisional rates with H$_2$ are taken from the Leiden Atomic and
Molecular Database (LAMDA; \citealt{Scho05}). The inclusion of these
coolants allows the thermal balance to be more accurately determined in warm
dense gas at high extinction. The treatment of H$_2$ self-shielding has also
been updated to use the results of detailed calculations performed by \citet{Lee96}. The chemical
network links 131 species in over 1700 gas-phase reactions; only
H$_{2}$ is formed by surface chemistry; freeze-out of species on to grain
surfaces is excluded. The UCL\_PDR code has been validated against all
other commonly used PDR codes \citep{Roel07}. The main elemental
abundances relative to hydrogen that have been adopted allow for some
depletion onto dust, and are C: $1.4\times 10^{-4} $; O: $3.2\times 10^{-4} $; N: $6.5\times 10^{-5} $; S:
$1.4\times 10^{-6} $; Mg: $5.1\times 10^{-6} $; and He: $7.5\times 10^{-2} $. The adopted gas:dust mass ratio is
100.

The code requires the adoption of several important parameters that define
the local physical conditions. These are (i) the gas number density, or
equivalently (since the temperature is determined by the code) the thermal
pressure; (ii) the FUV radiation flux impinging on the slab boundary; and
(iii) the cosmic ray induced ionisation rate, here assumed to be
independent of space and time. Most importantly, in this computation there is also an additional term: (iv) an assumed additional heating source in the
filaments, as indicated by \citet{Ferl09}. We discuss each of these
parameters in turn.

(i) In most cases the thermal pressure was taken to be $P = k_{B} (4
\times 10^{6}$ cm$^{-3}$ K), where $k_{B}$ is Boltzmann's constant. This
value of $P$ is close to the the average values measured by \citet{Sand04} for filaments showing optical line emission in NGC~1275 (see also
\citealt{Sand07}). The \citet{Sand04} measurements show a wide
scatter, and the average shows an upward gradient in pressure with
decreasing cluster radius. Therefore, in some computations we have set the
pressure to be one order of magnitude larger than stated above.

(ii) The incident FUV radiation field was taken to be $I$ times the
\citet{Habi68} radiation field mean intensity in the interstellar medium
of the Milky Way galaxy.
Values of $I$ of 10 and 100 were adopted, on the assumption that the
central galaxy in the cluster is, like NGC~1275, a
bright galaxy with a central AGN. Of course, obscuration of
stellar and AGN radiation affects the intensity impinging on the
filaments, and the nature and strength of the incident radiation
field are uncertain.

(iii) The cosmic ray induced ionisation rate, $\zeta$, was assigned values
in the range of $5 \times 10^{-17}$ s$^{-1}$ to $5 \times 10^{-13}$
s$^{-1}$; the higher values reflect the suggestion of \citet{Ferl09} for filaments in NGC~1275. Values around $5 \times 10^{-17}$
s$^{-1}$ are often assumed for clouds within several hundred parsecs of
the Sun and even in some studies of extragalactic regions.

(iv) A special feature and most important modification of the present
calculations is the inclusion of an
unspecified heat source, in addition to the familiar main heat sources of
photo-absorption and cosmic ray induced ionisation. The
additional heating mechanism, is here - following \citet{Ferl09} -
assumed to be dissipation. We incorporate it by adding a heat source
term, $H$, to the equation of thermal balance. It is time-independent and
uniform everywhere, and specifies the thermal energy added per unit time
per unit volume by processes not otherwise included in the code.
\citet{Ferl08, Ferl09} found that $H$ must be within an order of
magnitude of $10^{-19}$ erg cm$^{-3}$ s$^{-1}$ if dissipation powers the
observed optical emission regions. \citet{Ferl09} chose to make
their heat source term proportional to gas number density. However,
since the physics of the dissipation process is obscure, this choice is
arbitrary, and we have set $H$ to be independent of all parameters. The
relation between the terms is $H_{0}$(Ferland) = $H$(Bayet)/(n/1 cm$^{-3}$), or,
equivalently, $H_{0}$(Ferland) = $H$(Bayet)(T/4 x 10$^{6}$ K cm$^{-3}$) for the
assumed pressure.
Cosmic ray induced ionisation of a region with a hydrogen number density
of $10^{3}$ cm$^{-3}$ will be subject to a comparable heating rate (i.e. $H=10^{-19}$erg cm$^{-3}$ s$^{-1}$) due
to cosmic rays if $\zeta$ is within an order of magnitude of $10^{-12}$
s$^{-1}$.

\begin{landscape}
\begin{table}
  \caption{Summary of model runs. When the symbol for a molecular
  species is underlined, its fractional abundance is above
  1$\times10^{-10}$ otherwise it is simply above 1$\times 10^{-12}$
  (limit of detectability assumed). The temperature mentioned in the column 6 is self-consistently determined at the particular value of A$_{\rm v}$ shown.}\label{tab:obs}
  \vspace{-0.6cm}
  \hspace{-0.4cm}
  \resizebox{25.5cm}{!}{
    \begin{tabular}{c c c c c | c c l}
   & &  & & & & &\\
   Model & & & & & & &\\
   \hline
   A$_{\rm v}=3$ & $H$ & $P$ & $I$ & $\zeta$ & \emph{T} & CO/H & Interesting\\
   &(erg cm$^{-3}$ s$^{-1}$) & (cm$^{-3}$ K) & (Habing) & (s$^{-1}$) & (K) & & molecules\\
& &  & & & & &\\
  1&1$\times10^{-20}$ & 4$\times10^{6}$ & 100 & 5$\times10^{-14}$ & 2866.9 & 1.7$\times10^{-9}$ & \underline{O}, \underline{OH}, H$_{3}$$^{+}$, \underline{C$^{+}$}, CN, \underline{C}, \underline{H$_{2}$}, \underline{H}, H$_{2}$O, \underline{CO}\\
  2&1$\times10^{-22}$ & 4$\times10^{6}$ & 100 & 5$\times10^{-14}$ & 91.4 & 4.1$\times10^{-6}$ & \underline{H$_{3}$$^{+}$}, \underline{C$^{+}$}, \underline{CN}, \underline{C}, \underline{H$_{2}$}, \underline{H}, \underline{H$_{2}$O}, HCN, C$_{2}$, \underline{H$_{3}$O$^{+}$}, C$_{2}$H, \underline{HCO$^{+}$}, \underline{CO}, \underline{O}, \underline{OH}, CO$_{2}$, HNC\\
  3&1$\times10^{-20}$ & 4$\times10^{6}$ & 100 & 5$\times10^{-15}$ & 68.2   & 6.5$\times10^{-6}$ & \underline{H$_{3}$$^{+}$}, \underline{C$^{+}$}, \underline{CN}, \underline{C}, \underline{H$_{2}$}, \underline{H}, \underline{H$_{2}$O}, HCN, NH$_{3}$, \underline{C$_{2}$}, \underline{H$_{3}$O$^{+}$}, \underline{C$_{2}$H}, OCN, SO, \underline{CS}, \underline{HCO$^{+}$}, \underline{CO}, \underline{O}, \underline{OH}, H$_{2}$CO, CO$_{2}$, \underline{HNC}, HNO\\
  4&1$\times10^{-22}$ & 4$\times10^{6}$ & 100 & 5$\times10^{-15}$ & 51.3   & 3.5$\times10^{-6}$ & \underline{H$_{3}$$^{+}$}, \underline{C$^{+}$}, \underline{CN}, \underline{C}, \underline{H$_{2}$}, \underline{H}, \underline{H$_{2}$O}, HCN, NH$_{3}$, \underline{C$_{2}$}, \underline{H$_{3}$O$^{+}$}, \underline{C$_{2}$H}, OCN, SO, \underline{CS}, \underline{HCO$^{+}$}, \underline{CO}, \underline{O}, \underline{OH}, H$_{2}$CO, CO$_{2}$, HNC, HNO\\
  5&1$\times10^{-22}$ & 4$\times10^{6}$ & 100 & 5$\times10^{-17}$ & 25.3   & 1.2$\times10^{-5}$ & \underline{H$_{3}$$^{+}$}, \underline{C$^{+}$}, \underline{CN}, HCO, \underline{C}, \underline{H$_{2}$}, \underline{H}, \underline{H$_{2}$O}, \underline{HCN}, NH$_{3}$, \underline{C$_{2}$}, H$_{3}$O$^{+}$, C$_{2}$H, OCN, HCS, H$_{2}$CS, NS, \underline{SO}, \underline{CS}, HCO$^{+}$, \underline{CO},\\
  &&&&&&& \underline{O}, \underline{OH}, H$_{2}$CO, HS, \underline{CO$_{2}$}, \underline{HNC}, HNO\\
  6&1$\times10^{-22}$ & 4$\times10^{6}$ & 10 & 5$\times10^{-13}$ & 2479.6 & 9.9$\times10^{-11}$  & \underline{H$_{3}$$^{+}$}, \underline{C$^{+}$}, \underline{C}, \underline{H$_{2}$}, \underline{H}, H$_{2}$O, \underline{CO}, \underline{O}, \underline{OH}\\
  7&1$\times10^{-23}$ & 4$\times10^{6}$ & 10 & 5$\times10^{-13}$ & 2456.1 & 1.0$\times10^{-10}$  & \underline{H$_{3}$$^{+}$}, \underline{C$^{+}$}, \underline{C}, \underline{H$_{2}$}, \underline{H}, H$_{2}$O, \underline{CO}, \underline{O}, \underline{OH}\\
  8&1$\times10^{-20}$ & 4$\times10^{6}$ & 10 & 5$\times10^{-14}$ & 87.5  & 3.9$\times10^{-6}$ & \underline{C$^{+}$}, CN, \underline{C}, \underline{H$_{2}$}, \underline{H}, \underline{H$_{2}$O}, HCN, C$_{2}$, \underline{H$_{3}$O$^{+}$}, C$_{2}$H, \underline{HCO$^{+}$}, \underline{CO}, \underline{O}, \underline{OH}, CO$_{2}$, HNC\\
  9&1$\times10^{-22}$ & 4$\times10^{6}$ & 10 & 5$\times10^{-14}$ & 86.2  & 3.8$\times10^{-6}$ & \underline{H$_{3}$$^{+}$}, \underline{C$^{+}$}, CN, \underline{C}, \underline{H$_{2}$}, \underline{H}, \underline{H$_{2}$O}, HCN, C$_{2}$, \underline{H$_{3}$O$^{+}$}, C$_{2}$H, \underline{HCO$^{+}$}, \underline{CO}, \underline{O}, \underline{OH}, CO$_{2}$, HNC\\
  10&1$\times10^{-15}$-1$\times10^{-19}$ & 4$\times10^{6}$ & 10 & 5$\times10^{-15}$ & 10000.0  & 5.35$\times10^{-14}$ & \underline{C$^{+}$}, \underline{C}, \underline{H$_{2}$}, \underline{H}, \underline{O}, OH\\
  11&1$\times10^{-20}$ & 4$\times10^{6}$ & 10 & 5$\times10^{-15}$ & 61.8  & 5.8$\times10^{-6}$ & \underline{H$_{3}$$^{+}$}, \underline{C$^{+}$}, \underline{CN}, \underline{C}, \underline{H$_{2}$}, \underline{H}, \underline{H$_{2}$O}, HCN, NH$_{3}$, \underline{C$_{2}$}, \underline{H$_{3}$O$^{+}$}, \underline{C$_{2}$H}, OCN, SO, \underline{CS}, \underline{HCO$^{+}$}, \underline{CO}, \underline{O}, \underline{OH}, H$_{2}$CO, CO$_{2}$, \underline{HNC}, HNO\\
  12&1$\times10^{-22}$ & 4$\times10^{6}$ & 10 & 5$\times10^{-15}$ & 51.0  & 3.7$\times10^{-6}$ & \underline{H$_{3}$$^{+}$}, \underline{C$^{+}$}, \underline{CN}, \underline{C}, \underline{H$_{2}$}, \underline{H}, \underline{H$_{2}$O}, HCN, NH$_{3}$, \underline{C$_{2}$}, \underline{H$_{3}$O$^{+}$}, \underline{C$_{2}$H}, OCN, SO, \underline{CS}, \underline{HCO$^{+}$}, \underline{CO}, \underline{O}, \underline{OH}, H$_{2}$CO, CO$_{2}$, HNC, HNO\\
  13& 0 & 4$\times10^{6}$ & 10 & 5$\times10^{-15}$ & 50.8 & 3.7$\times10^{-6}$ & \underline{H$_{3}$$^{+}$}, \underline{C$^{+}$}, \underline{CN}, \underline{C}, \underline{H$_{2}$}, \underline{H}, \underline{H$_{2}$O}, HCN, NH$_{3}$, \underline{C$_{2}$}, \underline{H$_{3}$O$^{+}$}, \underline{C$_{2}$H}, OCN, SO, \underline{CS}, \underline{HCO$^{+}$}, \underline{CO}, \underline{O}, \underline{OH}, H$_{2}$CO, CO$_{2}$, HNC, HNO\\
  14&1$\times10^{-22}$ & 4$\times10^{6}$ & 10 & 5$\times10^{-17}$ & 13.8  & 1.4$\times10^{-5}$ & \underline{H$_{3}$$^{+}$}, \underline{C$^{+}$}, \underline{CN}, HCO, \underline{C}, \underline{H$_{2}$}, \underline{H}, \underline{H$_{2}$O}, \underline{HCN}, \underline{NH$_{3}$}, C$_{2}$, H$_{2}$S, H$_{3}$O$^{+}$, C$_{2}$H, OCS, \underline{OCN}, HCS, H$_{2}$CS, NS, \underline{SO}, \underline{CS}, \\
  &&&&&&& HCO$^{+}$, \underline{CO}, \underline{O}, \underline{OH}, H$_{2}$CO, HS, \underline{CO$_{2}$}, \underline{HNC}, \underline{HNO}\\
  15& 0 & 4$\times10^{6}$ & 10 & 5$\times10^{-17}$ & 12.4  & 1.4$\times10^{-5}$ & \underline{H$_{3}$$^{+}$}, \underline{C$^{+}$}, \underline{CN}, HCO, \underline{C}, \underline{H$_{2}$}, \underline{H}, \underline{H$_{2}$O}, \underline{HCN}, \underline{NH$_{3}$}, C$_{2}$, H$_{2}$S, H$_{3}$O$^{+}$, C$_{2}$H, OCS, \underline{OCN}, HCS, \underline{H$_{2}$CS}, NS, \underline{SO}, \underline{CS}, \\
  &&&&&&& HCO$^{+}$, \underline{CO}, \underline{O}, \underline{OH}, H$_{2}$CO, HS, \underline{CO$_{2}$}, \underline{HNC}, \underline{HNO}\\
  16&1$\times10^{-20}$ & 4$\times10^{7}$ & 10 & 5$\times10^{-15}$ & 43.8  & 1.3$\times10^{-5}$ & \underline{H$_{3}$$^{+}$}, \underline{C$^{+}$}, \underline{CN}, \underline{C}, \underline{H$_{2}$}, \underline{H}, \underline{H$_{2}$O}, \underline{HCN}, \underline{NH$_{3}$}, \underline{C$_{2}$}, H$_{2}$S, \underline{H$_{3}$O$^{+}$}, \underline{C$_{2}$H}, OCS, \underline{OCN}, HCS, H$_{2}$CS, NS, \underline{SO}, \underline{CS}, \\
  &&&&&&& \underline{HCO$^{+}$}, \underline{CO}, \underline{O}, \underline{OH}, \underline{H$_{2}$CO}, HS, \underline{CO$_{2}$}, \underline{HNC}, \underline{HNO}, C$_{2}$N, H$_{2}$CN \\
  17&1$\times10^{-22}$ & 4$\times10^{7}$ & 10 & 5$\times10^{-15}$ & 41.1 & 1.3$\times10^{-5}$ & \underline{H$_{3}$$^{+}$}, \underline{C$^{+}$}, \underline{CN}, \underline{C}, \underline{H$_{2}$}, \underline{H}, \underline{H$_{2}$O}, \underline{HCN}, \underline{NH$_{3}$}, \underline{C$_{2}$}, H$_{2}$S, \underline{H$_{3}$O$^{+}$}, \underline{C$_{2}$H}, OCS, \underline{OCN}, HCS, \underline{H$_{2}$CS}, NS, \underline{SO}, \underline{CS},\\
  &&&&&&& \underline{HCO$^{+}$}, \underline{CO}, \underline{O}, \underline{OH}, \underline{H$_{2}$CO}, HS, \underline{CO$_{2}$}, \underline{HNC}, \underline{HNO}, C$_{2}$N, H$_{2}$CN\\
18&1$\times10^{-20}$ & 4$\times10^{6}$ & 10 & 5$\times10^{-17}$ & 74.8 & 1.3$\times10^{-5}$ & \underline{H$_{3}$$^{+}$}, \underline{C$^{+}$}, \underline{CN}, \underline{C}, \underline{H$_{2}$}, \underline{H}, \underline{H$_{2}$O}, \underline{HCN}, \underline{NH$_{3}$}, \underline{C$_{2}$}, H$_{3}$O$^{+}$, C$_{2}$H, OCS, OCN, HCS, H$_{2}$CS, NS, \underline{SO}, CS,\\
   &&&&&&& \underline{HCO$^{+}$}, \underline{CO}, \underline{O}, \underline{OH}, \underline{H$_{2}$CO}, HS, \underline{CO$_{2}$}, \underline{HNC}, \underline{HNO},
   H$_{2}$CN\\
\hline
A$_{\rm v}=8$ & && & & & &\\
& & &  & & &&\\
   8&1$\times10^{-20}$ & 4$\times10^{6}$ & 10 & 5$\times10^{-14}$ & 101.7 & 4.9$\times10^{-6}$ & \underline{H$_{3}$$^{+}$}, \underline{C$^{+}$}, CN, \underline{C}, \underline{H$_{2}$}, \underline{H}, \underline{H$_{2}$O}, HCN, NH$_{3}$, C$_{2}$, \underline{H$_{3}$O$^{+}$}, C$_{2}$H, SO, \underline{HCO$^{+}$}, \underline{CO}, \underline{O}, \underline{OH}, CO$_{2}$, HNC\\
   9&1$\times10^{-22}$ & 4$\times10^{6}$ & 10 & 5$\times10^{-14}$ & 99.1 & 4.7$\times10^{-6}$ & \underline{H$_{3}$$^{+}$}, \underline{C$^{+}$}, CN, \underline{C}, \underline{H$_{2}$}, \underline{H}, \underline{H$_{2}$O}, HCN, NH$_{3}$, C$_{2}$, \underline{H$_{3}$O$^{+}$}, C$_{2}$H, SO, \underline{HCO$^{+}$}, \underline{CO}, \underline{O}, \underline{OH}, CO$_{2}$, HNC\\
   10&1$\times10^{-15}$-1$\times10^{-19}$ & 4$\times10^{6}$ & 10 & 5$\times10^{-15}$ & 10000.0 & 5.36$\times10^{-14}$ & \underline{C$^{+}$}, \underline{C}, \underline{H$_{2}$}, \underline{H}, \underline{O}, OH\\
   11&1$\times10^{-20}$ & 4$\times10^{6}$ & 10 & 5$\times10^{-15}$ & 78.7 & 8.3$\times10^{-6}$ & \underline{H$_{3}$$^{+}$}, \underline{C$^{+}$}, \underline{CN}, \underline{C}, \underline{H$_{2}$}, \underline{H}, \underline{H$_{2}$O}, HCN, \underline{NH$_{3}$}, \underline{C$_{2}$}, \underline{H$_{3}$O$^{+}$}, \underline{C$_{2}$H}, OCN, SO, \underline{CS}, \underline{HCO$^{+}$}, \underline{CO}, \underline{O}, \underline{OH}, H$_{2}$CO, CO$_{2}$, \underline{HNC}, HNO\\
   12&1$\times10^{-22}$ & 4$\times10^{6}$ & 10 & 5$\times10^{-15}$ & 60.2 & 5.6$\times10^{-6}$ & \underline{H$_{3}$$^{+}$}, \underline{C$^{+}$}, \underline{CN}, \underline{C}, \underline{H$_{2}$}, \underline{H}, \underline{H$_{2}$O}, HCN, NH$_{3}$, \underline{C$_{2}$}, \underline{H$_{3}$O$^{+}$}, \underline{C$_{2}$H}, OCN, SO, \underline{CS}, \underline{HCO$^{+}$}, \underline{CO}, \underline{O}, \underline{OH}, H$_{2}$CO, CO$_{2}$, \underline{HNC}, HNO\\
   13& 0 & 4$\times10^{6}$ & 10 & 5$\times10^{-15}$ & 59.9  & 5.5$\times10^{-6}$ & \underline{H$_{3}$$^{+}$}, \underline{C$^{+}$}, \underline{CN}, \underline{C}, \underline{H$_{2}$}, \underline{H}, \underline{H$_{2}$O}, HCN, NH$_{3}$, \underline{C$_{2}$}, \underline{H$_{3}$O$^{+}$}, \underline{C$_{2}$H}, OCN, SO, \underline{CS}, \underline{HCO$^{+}$}, \underline{CO}, \underline{O}, \underline{OH}, H$_{2}$CO, CO$_{2}$, \underline{HNC}, HNO\\
   14&1$\times10^{-22}$ & 4$\times10^{6}$ & 10 & 5$\times10^{-17}$ & 15.0  & 1.4$\times10^{-5}$ & \underline{H$_{3}$$^{+}$}, \underline{C$^{+}$}, \underline{CN}, HCO, \underline{C}, \underline{H$_{2}$}, \underline{H}, \underline{H$_{2}$O}, \underline{HCN}, \underline{NH$_{3}$}, C$_{2}$, H$_{2}$S, H$_{3}$O$^{+}$, C$_{2}$H, \underline{OCS}, \underline{OCN}, H$_{2}$CS, CH$_{3}$OH, NS, \underline{SO}, \underline{CS}, \\
  &&&&&&& HCO$^{+}$, \underline{CO}, \underline{O}, \underline{OH}, H$_{2}$CO, HS, \underline{CO$_{2}$}, \underline{HNC}, HNO\\
   15&0 & 4$\times10^{6}$ & 10 & 5$\times10^{-17}$ & 13.4  & 1.4$\times10^{-5}$ & \underline{H$_{3}$$^{+}$}, \underline{C$^{+}$}, \underline{CN}, HCO, \underline{C}, \underline{H$_{2}$}, \underline{H}, \underline{H$_{2}$O}, \underline{HCN}, \underline{NH$_{3}$}, C$_{2}$, H$_{2}$S, H$_{3}$O$^{+}$, C$_{2}$H, \underline{OCS}, \underline{OCN}, H$_{2}$CS, CH$_{3}$OH, NS, \underline{SO}, \underline{CS}, \\
  &&&&&&& HCO$^{+}$, \underline{CO}, \underline{O}, \underline{OH}, H$_{2}$CO, HS, \underline{CO$_{2}$}, \underline{HNC}, HNO\\
   16&1$\times10^{-20}$ & 4$\times10^{7}$ & 10 & 5$\times10^{-15}$ & 47.5 & 1.3$\times10^{-5}$ & \underline{H$_{3}$$^{+}$}, \underline{C$^{+}$}, \underline{CN}, \underline{C}, \underline{H$_{2}$}, \underline{H}, \underline{H$_{2}$O}, \underline{HCN}, \underline{NH$_{3}$}, \underline{C$_{2}$}, H$_{2}$S, \underline{H$_{3}$O$^{+}$}, \underline{C$_{2}$H}, OCS, \underline{OCN}, HCS, \underline{H$_{2}$CS}, NS, \underline{SO}, \underline{CS},\\
   &&&&&&& \underline{HCO$^{+}$}, \underline{CO}, \underline{O}, \underline{OH}, \underline{H$_{2}$CO}, HS, \underline{CO$_{2}$}, \underline{HNC}, \underline{HNO},
   C$_{2}$N\\
   17&1$\times10^{-22}$ & 4$\times10^{7}$ & 10 & 5$\times10^{-15}$ & 45.2 & 1.3$\times10^{-5}$ & \underline{H$_{3}$$^{+}$}, \underline{C$^{+}$}, \underline{CN}, \underline{C}, \underline{H$_{2}$}, \underline{H}, \underline{H$_{2}$O}, \underline{HCN}, \underline{NH$_{3}$}, \underline{C$_{2}$}, H$_{2}$S, \underline{H$_{3}$O$^{+}$}, \underline{C$_{2}$H}, OCS, \underline{OCN}, HCS, \underline{H$_{2}$CS}, NS, \underline{SO}, \underline{CS},\\
   &&&&&&& \underline{HCO$^{+}$}, \underline{CO}, \underline{O}, \underline{OH}, \underline{H$_{2}$CO}, HS, \underline{CO$_{2}$}, \underline{HNC}, \underline{HNO},
   C$_{2}$N\\
   18&1$\times10^{-20}$ & 4$\times10^{6}$ & 10 & 5$\times10^{-17}$ & 63.3 & 1.4$\times10^{-5}$ & \underline{H$_{3}$$^{+}$}, \underline{C$^{+}$}, \underline{CN}, \underline{C}, \underline{H$_{2}$}, \underline{H}, \underline{H$_{2}$O}, \underline{HCN}, \underline{NH$_{3}$}, C$_{2}$, H$_{2}$S, H$_{3}$O$^{+}$, C$_{2}$H, OCS, \underline{OCN},H$_{2}$CS, NS, \underline{SO}, \underline{CS},\\
   &&&&&&& \underline{HCO$^{+}$}, \underline{CO}, \underline{O}, \underline{OH}, H$_{2}$CO, HS, \underline{CO$_{2}$}, \underline{HNC}, \underline{HNO}\\
\end{tabular}}
\end{table}
\end{landscape}

The higher values of our assumed heating rates per
unit volume and cosmic ray induced ionisation rates roughly
approach the values found by \citet{Ferl08, Ferl09} to be
compatible with the optical emission line spectra in NGC~1275.
However, we limited the upper range of the assumed rates to
avoid heating the gas beyond the maximum temperatures at which
molecules are abundant. This limiting is consistent with the
values of $H$ and $\zeta$ decreasing with depth into a filament.

\section{Results}\label{sec:resu}

The code determines self-consistently the steady-state thermal and
chemical properties of a constant-pressure filament subjected to
the three heating processes described in Section \ref{sec:model};
of these, $H$ and $\zeta$ affect the bulk of the filament. In
contrast, the effects of $I$ are confined to narrow surface
regions, as a comparison of results given in Figures \ref{fig:2}
and \ref{fig:3} for our models 14 and 5 (which differ only in $I$)
confirms. The thermal and chemical properties are computed as
functions of A$_{\rm v}$, measured from the outer edge of the
filament. We present in Table \ref{tab:obs} a summary of the
models explored and the corresponding thermal and chemical
properties at two values of A$_{\rm v}$, 3 and 8 mag. These values
are intended to represent conditions near the edge and in the dark
interior of a filament. We do not list results at A$_{\rm v} = 8$
for Models 1-5 as they should not differ from those at A$_{\rm
v}=3$. We include in the final column of Table 1 a list of
molecules that are predicted to be fairly abundant in each model
(underlined species) and molecules that may be near the current
limit of detection (non-underlined species). Evidently, some
models are expected to be chemically very rich, while others
should be chemically poor. Table \ref{tab:obs2} contains values
at A$_{\rm v} = 8$ of the fractional abundances of the species. In
this section we focus on the results shown in various figures, but
the trends that we infer from those figures are also seen in the
results in Table \ref{tab:obs2}.

\begin{figure*}
  \centering
  \includegraphics[height=18cm]{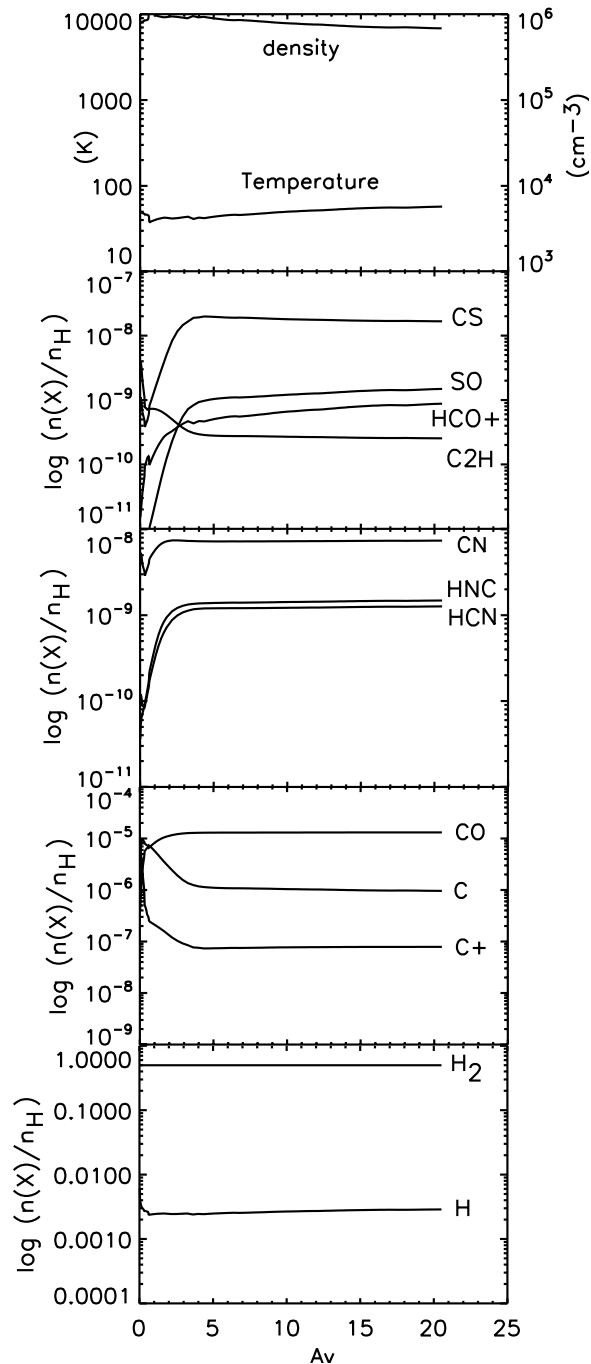}
  \caption{Fractional abundances with respect to total hydrogen number
  density of various species (X) as functions of optical depth
  (A$_{\rm v}$), derived from model 16. For each model presented in Figs \ref{fig:1}, \ref{fig:2} and \ref{fig:3},
  the plots on the top correspond to the density and the temperature
  profiles. The chemical formula of each molecular species studied is
  indicated on the plots. When not mentioned, it means that the
  fractional abundance of X is lower than the arbitrary limit
  of detectability assumed to be equal to $1 \times 10^{-12}$.
  We have used solid lines for representing the
  model having an additional heating source of $1 \times
  10^{-20}$ erg cm$^{-3}$ s$^{-1}$ and the dashed lines show the
  models having an additional heating source of $1 \times
  10^{-22}$ erg cm$^{-3}$ s$^{-1}$.}\label{fig:1}
\end{figure*}

 \begin{figure*}
  \centering
  \includegraphics[height=18cm]{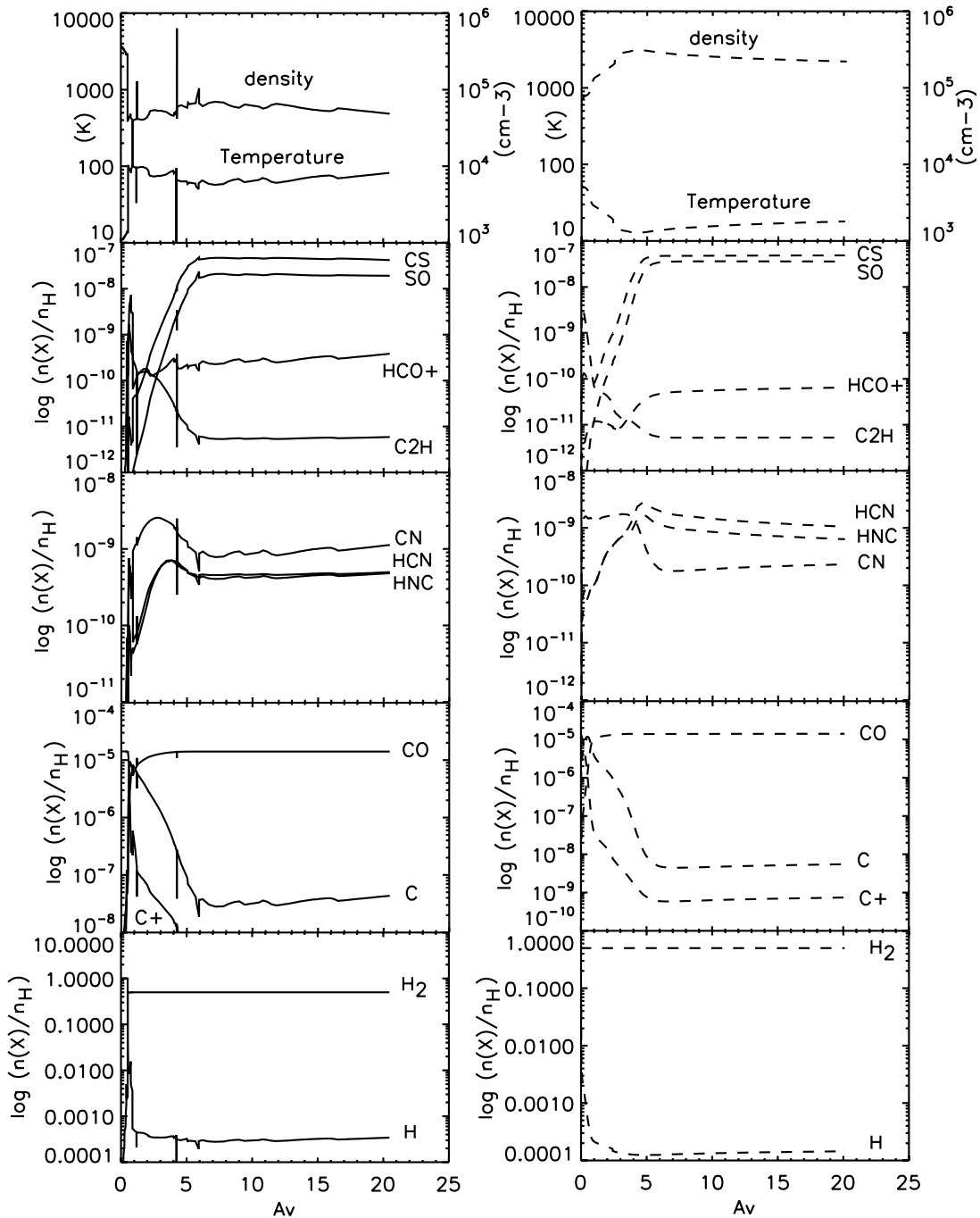}
  \caption{Fractional abundances with respect to total hydrogen number
  density of various species (X) as functions of optical depth
  (A$_{\rm v}$), derived from Model 18 (left) and Model 14
  (right). We have used solid lines for representing the
  models having an additional heating source of $1 \times
  10^{-20}$ erg cm$^{-3}$ s$^{-1}$ (i.e Model 18) and,
  as previously, the dashed lines show the
  models having an additional heating source of $1 \times
  10^{-22}$ erg cm$^{-3}$ s$^{-1}$ (i.e Model 14).
  See caption of Fig.\ref{fig:1}.}\label{fig:2}
\end{figure*}

 \begin{figure*}
  \centering
  \includegraphics[height=18cm]{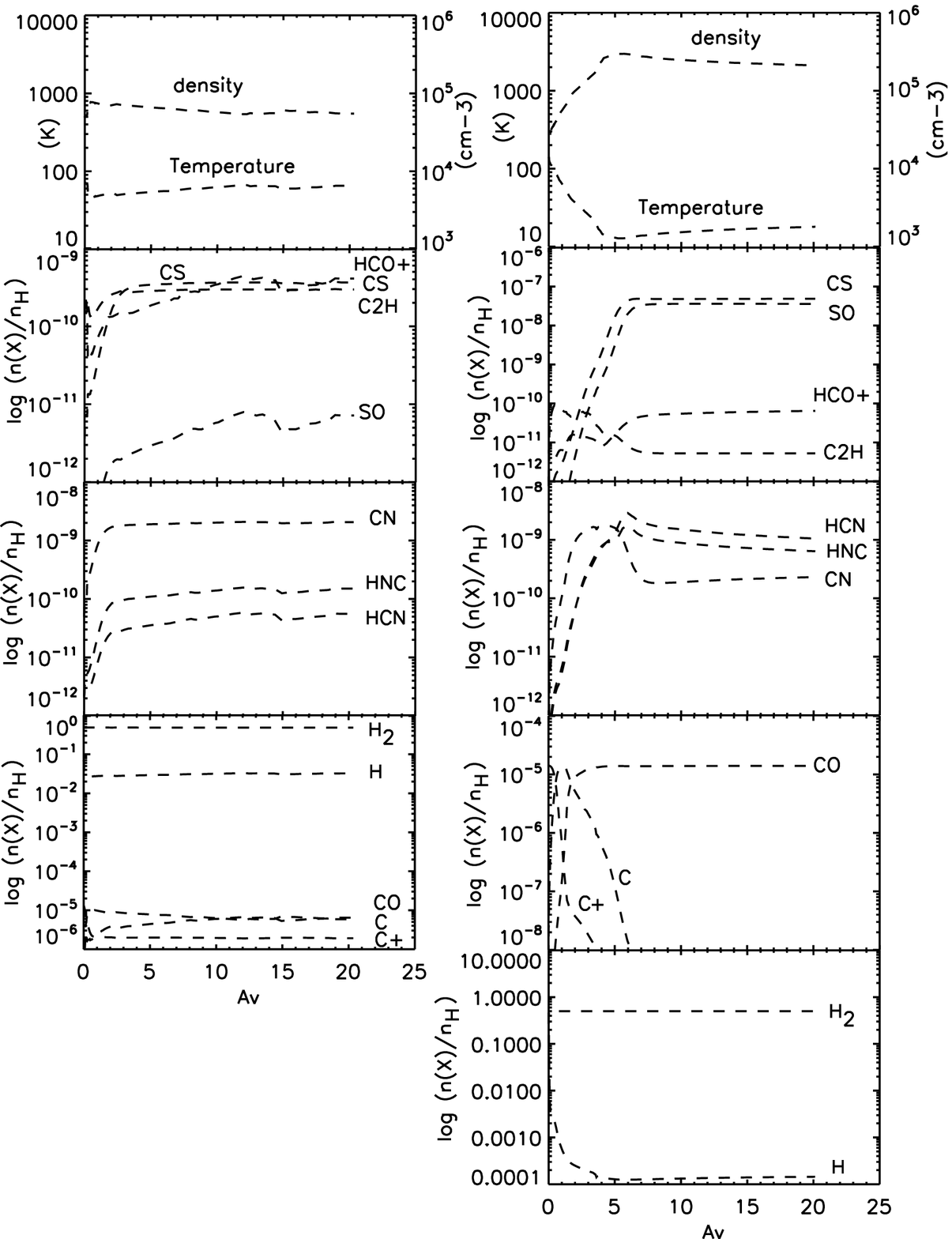}
  \caption{Fractional abundances with respect to total hydrogen number
  density of various species (X) as functions of optical depth
  (A$_{\rm v}$), derived from Model 12 (left) and Model 5
  (right). As previously, the dashed lines show the
  models having an additional heating source of $1 \times
  10^{-22}$ erg cm$^{-3}$ s$^{-1}$ (i.e both Models 12 and 5). Here the difference between the models is the cosmic ray ionisation rate used ($\zeta=5 \times
  10^{-15}$ s$^{-1}$ for Model 12 whereas Model 5 has $\zeta=5 \times
  10^{-17}$ s$^{-1}$) and the FUV radiation field ($I=10$ Habing for Model 12 whereas Model 5 shows $I=100$ Habing). See caption of Fig.\ref{fig:1} and text in Sect. \ref{sec:resu}.}\label{fig:3}
\end{figure*}

The chemistry varies little with  A$_{\rm v}$ as A$_{\rm v}$ increases beyond 8. However, the figures that we show give results to higher values of A$_{\rm v}$ because the visual extinctions of the clumps of which a filament is composed may range to higher values. The maximum A$_{\rm v}$ from its edge to its centre is probably that of the maximum mass isothermal sphere that is stable against self-gravity. For clumps with dust and elemental abundances like those of of clouds in the local interstellar medium, this A$_{\rm v}$ is given roughly by

\begin{equation}
A_{v_{max}} = 15 \rm{mag} [\emph{n}(H_{2})\emph{T}/(4\times 10^6 \rm{ K} \rm{cm}^{-3})]^{1/2}
\end{equation}

This result is based on the expression for the maximum external pressure that can be applied to a Bonner-Ebert sphere without causing it to collapse, the assumption that the pressure in the sphere is roughly constant, and that a column density of $2 \times 10^{21}$ cm$^{-2}$ hydrogen nuclei corresponds to one magnitude of visual extinction.

The temperature quoted in Table \ref{tab:obs} is the kinetic temperature determined self-consistently with the chemistry, given for depths into the filament corresponding to these two values of A$_{\rm v}$. The fractional abundance of CO determined by the code at those same depths is also given, as a crude measure of the effectiveness of the chemistry in the filament. It ranges from a value of $1.3\times 10^{-5}$, corresponding to a rich chemistry in which about 10\% of available carbon is in CO, to very small values. We summarise in Table \ref{tab:obs} the chemistry determined for each model by simply listing molecules familiar from studies of galactic interstellar media; more species than those listed are present. Molecules listed have a fractional abundances of at least $1\times 10^{-12}$ (adopted as an arbitrary limit of detectability), and those underlined have fractional abundances of at least $1\times 10^{-10}$. Detections of C$_{2}$ and CO$_{2}$ will be unlikely, but these molecules are listed for completeness. More information about the dependence of gas density, temperature and chemistry on A$_{\rm v}$ for particular models is found in Figures \ref{fig:1}, \ref{fig:2} and \ref{fig:3}, described below.

\vspace*{0.3cm}
\textit{Variations in the additional heating rate:}

Model 10 explores the effect of varying the additional rate, $H$, over the range $1\times 10^{-15}$ to $1\times 10^{-19}$ erg cm$^{-3}$ s$^{-1}$. The high values of $H$ dominate other heating mechanisms and overwhelm the cooling mechanisms. The resulting temperature is so high that very few molecules can survive. Since $H$ is assumed to be independent of depth, this result is true at all depths.

For values of $H$ of $1\times 10^{-20}$ erg cm$^{-3}$ s$^{-1}$ and smaller, a rich chemistry can exist but is sensitive to the adopted cosmic ray ionisation rate. For example, Models 11 ($H = 1\times 10^{-20}$ erg cm$^{-3}$ s$^{-1}$) and 12 ($H = 1\times 10^{-22}$ erg cm$^{-3}$ s$^{-1}$) are seen in Table \ref{tab:obs} to have almost identical extensive chemistries, while Model 13 ($H = 0$) is indistinguishable from Model 12. Thus, $H = 1\times 10^{-20}$ erg cm$^{-3}$ s$^{-1}$ appears to be a critical value; above this value, the heating is so great that the chemistry is largely suppressed, while below this value a rich chemistry occurs and is insensitive to $H$. However, as mentioned above, the value of $\zeta$ affects the richness of the chemistry for a particular value of $H$. For example, Models 11 and 8 (both with $H = 1\times 10^{-20}$ erg cm$^{-3}$ s$^{-1}$) differ in $\zeta$ ($5\times 10^{-15}$ s$^{-1}$ and $5\times 10^{-14}$ s$^{-1}$, respectively); Model 11 is distinctly richer in chemistry.

For heating due to dissipation to exceed that due to cosmic ray induced ionisation $H$ must exceed a value given roughly by

\begin{equation}
H_{min} =  1 \times 10^{-22} \rm{erg cm}^{-3}\rm{s}^{-1} \times (\emph{n}(H_{2})\emph{T})/(4 \times 10^6 \rm{K} \rm{cm}^{-3}) \times
\end{equation}
$$
(T/100\rm{K})^{-1} \times \zeta/(5 \times 10^{-17} \rm{s}^{-1}).
$$

This result is based on the assumption that one ionisation produces about 40 eV of thermal energy.

Figure \ref{fig:1} shows the dependences of the total number density and temperature  in the gas, and the chemical abundances, as functions of depth for Model 16. This model may have plausible values of the physical parameters. It shows a modest temperature and a rich chemistry. Evidently, the familar galactic tracers CS, SO, HCO$^{+}$, C$_{2}$H, CN, HNC and HCN should be useful for determining the nature of the intracluster material in cases where the parameters are similar to those of Model 16.

Figure \ref{fig:2} compares the chemistries predicted by two Models, 14 and 18, which both have low values of $\zeta$ but differ strongly in $H$ ($1\times 10^{-22}$ erg cm$^{-3}$ s$^{-1}$ and $1\times 10^{-20}$ erg cm$^{-3}$ s$^{-1}$, respectively). Some spikes are seen for Model 18 because the heating is only just matched by the cooling, and the balance is thus computationally delicate. Both models show a rich chemistry, but there are some physical and chemical differences. For example, the important tracer ion HCO$^{+}$ is more abundant in Model 18; and the molecules CN, HCN and HNC have a different priority order of abundance in the two models, caused by the higher temperature in Model 18.

\vspace*{0.3cm}
\textit{Variations in the cosmic ray ionisation rate:}

The value of $\zeta$ is an important parameter both for the chemistry and for the thermal balance. The cases with highest $\zeta$ examined here are Models 6 and 7, with $\zeta = 5\times 10^{-13}$s$^{-1}$. The value of $H$ is low in both Models ($1\times 10^{-22}$ erg cm$^{-3}$ s$^{-1}$ and $1\times 10^{-23}$ erg cm$^{-3}$ s$^{-1}$, respectively), so the major heat input is from the cosmic rays. The temperature is high and the chemistry is largely suppressed; even CO is held down to a fractional abundance of about $1\times 10^{-10}$. The high $\zeta$ but low CO abundance leads to a significant H$_{3}^{+}$ abundance. This suppression of the chemistry for high $\zeta$ is in agreement with the findings of \citet{Lepp96}(see also \citealt{Lepp98}).

Figure \ref{fig:3} compares results from Models 5 and 12 which both have low values of $H$ ($1\times 10^{-22}$ erg cm$^{-3}$ s$^{-1}$), but strongly differing values of $\zeta$ ($5\times 10^{-17}$ s$^{-1}$ and $5\times 10^{-15}$ s$^{-1}$, respectively). Both are chemically rich (Model 5 is slightly richer) but there are significant differences. For example, Model 5 has very much higher SO and CS abundances than Model 12, while the reversed is true for C$_{2}$H and HCO$^{+}$. The abundance order of priority of CN, HNC and HCN is reverse between Models 5 and 12, with the abundance of CN being one order of magnitude larger in the high $\zeta$ case, Model 12. This case also has abundant atomic species H, C and C$^{+}$ relative to Model 5. It is well known that the H:H$_{2}$ ratio, maintained in this case by the high $\zeta$, has a significant effect on the chemistry \citep{Rawl02}.

\vspace*{0.3cm}
\textit{Other variations:}

The adopted pressure, $P$, has a limited effect on the thermal and chemical behaviour. Comparing Models 11 and 16, or 12 and 17 (which differ only in $P$) shows that the temperature is slightly lower at higher pressure (as expected since the cooling is more effective), and that this slight difference generates a difference in the chemistry, with the cooler models being very slightly chemically richer.

The ambient radiation field can be an important source of heat at the cloud edge, comparable to $H = 1\times 10^{-20}$ erg cm$^{-3}$ s$^{-1}$ in unshielded situations. However, this heating effect decays rapidly with depth into the cloud. For the higher values of $I$ explored here, $I = 100$, the radiation field may still be making a contribution to the heating, even at A$_{\rm v} = 3$. We show in Table \ref{tab:obs3} a comparison of the heating rates from the main heating sources in several models at both low and high A$_{\rm v}$. Evidently, the cosmic rays and the dissipation (i.e. $H$) are the most important sources. Comparing Models 4 and 12 shows that the larger value of $I$ in Model 4 can be making no difference when $\zeta = 5\times 10^{-15}$ s$^{-1}$, because cosmic rays dominate the heating in these cases. However, comparing Models 5 and 14, both of which have a low value of $\zeta$, $5\times 10^{-17}$ s$^{-1}$, we see that the higher value of $I$ in Model 5 increases the temperature relative to Model 12 - though not enough to make a difference to the chemistry.

At low A$_{\rm v}$, Model 1 shows a surprisingly high temperature and poor chemistry compared to a similar model, 8, with lower value of $I$. Evidently, the radiation field can have a significant effect in the boundary layers of filaments. This comparison emphasises the ``flip-flop'' effect; a rich chemistry provides many coolants that generate a low temperature, enhancing the chemistry; and vice-versa. Model 1 is expected to show a much lower temperature at greater depths.

\begin{table*}
  \caption{Fractional abundances with respect to the total number of H-nuclei for models 8-18 at A$_{\rm v}=8$ mag. The power is indicated in parenthesis such as a(b)=a$\times 10^{b}$. When a dash is used, it means that the fractional abundance is below our assumed limit of detectability of 1$\times 10^{-12}$.}\label{tab:obs2}
  \hspace{-1.0cm}
  \begin{tabular}{c | c | c | c | c | c | c | c | c | c | c | c}
   Model: &  8 &  9 &  10 &  11 & 12 &  13 & 14 & 15 & 16 & 17 & 18\\
   \hline
   Species & & & & & & & & & & &\\
   H$_{3}^{+}$ & 2.84(-8) & 2.76(-8)& - & 5.36(-8)& 3.19(-8)& 3.16(-8)& 9.09(-10)& 8.35(-10)& 1.34(-8)& 1.27(-8)&2.74(-9)\\
   C$^{+}$ & 6.11(-6)& 6.18(-6)& 1.40(-5)& 1.70(-6)& 1.96(-6)& 1.97(-6)& 6.01(-10)& 5.32(-10)& 7.58(-8)& 7.43(-8)&3.27(-9)\\
   CN & 9.56(-11)& 9.33(-11)& - & 2.28(-9)& 1.99(-9)& 1.99(-9)& 1.81(-10)& 1.53(-10)& 7.24(-9)& 7.23(-9)&8.01(-10)\\
   HCO & - & - & - & -& -& -& 2.61(-12)& 2.56(-12)& -& -&-\\
   C & 3.23(-6)& 3.36(-6)& 2.06(-7)& 4.19(-6)& 6.66(-6)& 6.73(-6)& 4.48(-9)& 4.05(-9)& 1.07(-6)& 1.08(-6)&3.15(-8)\\
   H$_{2}$O & 2.30(-7)& 2.18(-7)& - & 9.90(-8)& 5.29(-8)& 5.22(-8)& 4.97(-7)& 5.14(-7)& 1.81(-7)& 1.78(-7)&3.82(-7)\\
   HCN & 2.87(-12)& 2.68(-12)& - & 9.25(-11)& 4.58(-11)& 4.51(-11)& 1.60(-9)& 2.02(-9)& 1.21(-9)& 1.20(-9)&4.61(-10)\\
   NH$_{3}$ & 1.22(-12)& 1.13(-12)& - & 1.15(-10)& 4.50(-11)& 4.42(-11)& 5.46(-8)& 6.85(-8)& 3.12(-9)& 3.02(-9)&1.66(-8)\\
   H$_{2}$S & - & - & - & - & -& -& 3.70(-11)& 4.07(-11)& 1.54(-12)& 1.60(-12)&1.01(-11)\\
   H$_{3}$O$^{+}$ & 3.14(-9)& 2.94(-9)& - & 8.10(-10)& 4.23(-10)& 4.16(-10)& 2.41(-11)& 2.33(-11)& 1.89(-10)& 1.77(-10)&5.11(-11)\\
   C$_{2}$H & 1.98(-12)& 2.00(-12)& - & 2.84(-10)& 2.95(-10)& 2.95(-10)& 5.21(-12)& 5.11(-12)& 2.74(-10)& 2.76(-10)&3.43(-12)\\
   OCS & - & - & - & -& -& -& 1.21(-10)& 1.50(-10)& 2.05(-11)& 2.19(-11)&1.125(-11)\\
   OCN & - & - & - & 8.66(-12) & 3.80(-12)& 3.74(-12)& 2.54(-8)& 3.26(-8)& 4.82(-10)& 5.07(-10)&3.52(-10)\\
   HCS& - & - & - & - & -& -& -& -& 3.55(-11)& 3.74(-11)&-\\
   H$_{2}$CS & - & - & - & - & -& -& 6.82(-11)& 6.72(-11)& 2.01(-10)& 2.13(-10)&4.92(-11)\\
   CH$_{3}$OH & - & - & - & - & -& -& 1.99(-12)& 2.44(-12) & -& -&-\\
   NS & - & - & - & - & -& -& 3.54(-12)& 2.96(-12)& 5.64(-12)& 5.42(-12)&1.13(-11)\\
   SO & 1.21(-12)& 1.08(-12)& - & 2.03(-11)& 4.84(-12)& 4.70(-12)& 3.60(-8)& 3.56(-8)& 1.14(-9)& 1.085(-9)&2.03(-8)\\
   CS & - & - & - & 3.65(-10)& 3.59(-10)& 3.58(-10)& 4.76(-8)& 4.67(-8)& 1.86(-8)& 1.93(-8)&4.64(-8)\\
   HCO$^{+}$ & 4.05(-10)& 3.72(-10)& - & 8.80(-10)& 2.94(-10)& 2.87(-10)& 5.32(-11)& 4.75(-11)& 5.94(-10)& 5.39(-10)&2.62(-10)\\
   CO & 4.86(-6)& 4.66(-6)& - & 8.31(-6)& 5.57(-6)& 5.50(-6)& 1.39(-5)& 1.38(-5)& 1.30(-5)& 1.30(-5)&1.41(-5)\\
   O & 2.53(-5)& 2.56(-5)& 3.18(-5)& 2.29(-5)& 2.60(-5)& 2.60(-5)& 9.13(-6)& 9.25(-6)& 1.73(-5)& 1.73(-5)&1.18(-5)\\
   OH & 1.48(-6)& 1.40(-6)& 1.67(-12)& 4.08(-7)& 2.23(-7)& 2.20(-7)& 2.58(-8)& 2.59(-8)& 1.17(-7)& 1.10(-7)&3.77(-8)\\
   H$_{2}$CO & - & - & - & 2.93(-11)& 2.23(-11)& 2.21(-11)& 5.13(-11)& 5.06(-11)& 5.18(-10)& 5.25(-10)&3.92(-11)\\
   HS & - & - & - & - & - & - & 1.23(-11) & 1.21(-11) & 8.89(-12) & 8.78(-12) &1.82(-11)\\
   HNC & 1.22(-11)& 1.15(-11)& - & 2.27(-10)& 1.31(-10)& 1.29(-10)&9.33(-10) & 1.15(-9)& 1.41(-9)& 1.40(-9)&4.22(-10)\\
   HNO & - & - & - & 6.61(-11)& 4.93(-11)& 4.88(-11)& 4.22(-10)& 2.84(-10)& 1.99(-9)& 1.95(-9)&7.43(-10)\\
   C$_{2}$N & - & - & - & - & - & -& -& -& 1.77(-12)& 1.75(-12)&-\\
   \hline
\end{tabular}
\end{table*}

\begin{table}
  \caption{Main contributors to the heating rates (erg s$^{-1}$ cm$^{-3}$) in several models (See Table \ref{tab:obs}).}\label{tab:obs3}
 \begin{tabular}{cccc}
  \hline
   A$_{\rm v}=3$ mag & & &\\
   heating sources & Model 1 & Model 8 & Model 18\\
   FUV & $8.8 \times 10^{-24}$& $8.8 \times 10^{-23}$& $1.7 \times 10^{-23}$\\
   $H$ & $1.0 \times 10^{-20}$& $1.0 \times 10^{-20}$& $1.0 \times 10^{-20}$\\
   Cosmic rays & $1.0 \times 10^{-21}$ & $3.4 \times 10^{-20}$& $4.0 \times 10^{-23}$\\
   \hline
   A$_{\rm v}=8$ mag & & &\\
   heating sources & Model 14 & Model 8 & Model 18\\
   FUV & $8.8 \times 10^{-26}$& $7.6 \times 10^{-27}$&$1.3 \times 10^{-27}$ \\
   $H$ & $1.0 \times 10^{-22}$& $1.0 \times 10^{-20}$& $1.0 \times 10^{-20}$\\
   Cosmic rays & $2.0 \times 10^{-22}$& $2.9 \times 10^{-20}$& $4.7 \times 10^{-23}$\\
   \hline
\end{tabular}
\end{table}


\section{Conclusions}\label{sec:conclu}

The self-consistent treatment of thermal and chemical behaviour
tends to lead to one of two alternative regimes: low temperature
and rich chemistry, or high temperature and poor chemistry. The
rich chemistry provides efficient cooling, enhancing the
chemistry, while elevated temperatures suppress the chemistry and
destroy coolants.

The parameter regime that is covered in this study shows that
there are two main contributors to the heating, the additional
source $H$ and cosmic rays. The ambient radiation field is
unlikely to be dominant and is in any case confined to a
skin-effect of a few visual magnitudes into the filament. The
density of the filament has been assumed in our calculations to be
uniform. Of course, is the gas is clumpy, then the radiation
fields may play a larger role at greater depths than in the
homogeneous case. Varying the pressure from the adopted value has
only a minor effect on the chemistry.

Very high values of $H$, $\geq 1\times 10^{-20}$ erg cm$^{-3}$
s$^{-1}$, tend to suppress the chemistry and consequent coolants,
so temperatures approach those of HII regions. Very high cosmic
ray ionisation rates, $\geq 1\times 10^{-13}$ s$^{-1}$, have a
similar effect. For $H \lesssim 1\times 10^{-20}$ erg cm$^{-3}$
s$^{-1}$ and $\zeta \lesssim 1\times 10^{-14}$ s$^{-1}$, the
heating and consequent chemical effects of the two processes
combine.

These rates considerably exceed the typical rates in dark clouds
near to the Sun (and may be even higher in the filamentary optical
regions). If these are representative of the conditions in
filamentary molecular regions, then there are many molecular
species that should be abundant and useful as tracers of the
physical conditions. CS, CN and HNC decline less in abundance with
increasing heating and ionisation rates than many species.
C$_{2}$H and HCO$^{+}$ increase in abundance with cosmic ray
ionisation rate for a wide range of rates but become less abundant
for very high rates. The ratio of those two abundances decreases
with increasing dissipative heating rate. N$_{2}$H$^{+}$, a
molecule used successfully to trace dense molecular clouds in the
Galaxy, is notable by its absence from our predictions.

In our recommendations for molecular line searches in cluster
filaments, we have been guided by the molecular abundances
predicted by our computations and by their sensitivity to changes
in heating and ionisation rates. The construction of
detailed models will be possible once more observational data
exist. However, we can give  some indication of the expected range
of HCO$^{+}$, C$_{2}$H and CN line intensities, assuming no beam
dilution. To do so, we have used
RADEX\footnote{http://www.sron.rug.nl/~vdtak/radex/radex.php}
developed by \citet{VanderTak07}, for plane-parallel geometry, for
Models 12, 14, 16 and 18 seen in Figs. \ref{fig:2} and
\ref{fig:3}. We have obtained for the HCO$^{+}$(1-0) transition,
values ranging from 0.62 K kms$^{-1}$ (Model 14) to 5.6 K
kms$^{-1}$ (Model 16). For the HCO$^{+}$(5-4) line, values range
from 0.2 K kms$^{-1}$ (Models 12 and 18) to 4.2 K kms$^{-1}$
(Model 16). For the CN(1-0) line, we have obtained values from 0.2
K kms$^{-1}$ (Model 14) to 7.6 K kms$^{-1}$ (Model 16). For the
CN(4-3) lines, we have obtained values ranging from 1.4$\times
10^{-2}$ K kms$^{-1}$ (Model 18) to 1.6 K kms$^{-1}$ (Model 16).

\bibliographystyle{aa}
\bibliography{references}

\begin{thebibliography}{29}
\expandafter\ifx\csname natexlab\endcsname\relax\def\natexlab#1{#1}\fi

\bibitem[{{Bayet} {et~al.}(2009){Bayet}, {Viti}, {Williams}, {Rawlings}, \&
  {Bell}}]{Baye09a}
{Bayet}, E., {Viti}, S., {Williams}, D.~A., {Rawlings}, J.~M.~C., \& {Bell}, T.
  2009, \apj, 696, 1466

\bibitem[{{Bell} {et~al.}(2007){Bell}, {Viti}, \& {Williams}}]{Bell07}
{Bell}, T.~A., {Viti}, S., \& {Williams}, D.~A. 2007, ArXiv e-prints, 704

\bibitem[{{Bell} {et~al.}(2005){Bell}, {Viti}, {Williams}, {Crawford}, \&
  {Price}}]{Bell05}
{Bell}, T.~A., {Viti}, S., {Williams}, D.~A., {Crawford}, I.~A., \& {Price},
  R.~J. 2005, \mnras, 357, 961

\bibitem[{{Conselice} {et~al.}(2001){Conselice}, {Gallagher}, \&
  {Wyse}}]{Cons01}
{Conselice}, C.~J., {Gallagher}, III, J.~S., \& {Wyse}, R.~F.~G. 2001, \aj,
  122, 2281

\bibitem[{{Crawford} {et~al.}(1999){Crawford}, {Allen}, {Ebeling}, {Edge}, \&
  {Fabian}}]{Craw99}
{Crawford}, C.~S., {Allen}, S.~W., {Ebeling}, H., {Edge}, A.~C., \& {Fabian},
  A.~C. 1999, \mnras, 306, 857

\bibitem[{{Crawford} \& {Fabian}(1992)}]{Craw92}
{Crawford}, C.~S. \& {Fabian}, A.~C. 1992, \mnras, 259, 265

\bibitem[{{de Jong} {et~al.}(1980){de Jong}, {Boland}, \& {Dalgarno}}]{DeJo80}
{de Jong}, T., {Boland}, W., \& {Dalgarno}, A. 1980, \aap, 91, 68

\bibitem[{{Ferland} {et~al.}(2008){Ferland}, {Fabian}, {Hatch}, {Johnstone},
  {Porter}, {van Hoof}, \& {Williams}}]{Ferl08}
{Ferland}, G.~J., {Fabian}, A.~C., {Hatch}, N.~A., {et~al.} 2008, \mnras, 386,
  L72

\bibitem[{{Ferland} {et~al.}(2009){Ferland}, {Fabian}, {Hatch}, {Johnstone},
  {Porter}, {van Hoof}, \& {Williams}}]{Ferl09}
{Ferland}, G.~J., {Fabian}, A.~C., {Hatch}, N.~A., {et~al.} 2009, \mnras, 392,
  1475

\bibitem[{{Habing}(1968)}]{Habi68}
{Habing}, H.~J. 1968, \bain, 19, 421

\bibitem[{{Ho} {et~al.}(2009){Ho}, {Lim}, \& {Dinh-V-Trung}}]{Ho09}
{Ho}, I.-T., {Lim}, J., \& {Dinh-V-Trung}. 2009, \apj, 698, 1191

\bibitem[{{Jaffe} {et~al.}(2005){Jaffe}, {Bremer}, \& {Baker}}]{Jaff05}
{Jaffe}, W., {Bremer}, M.~N., \& {Baker}, K. 2005, \mnras, 360, 748

\bibitem[{{Johnstone} {et~al.}(2007){Johnstone}, {Hatch}, {Ferland}, {Fabian},
  {Crawford}, \& {Wilman}}]{John07}
{Johnstone}, R.~M., {Hatch}, N.~A., {Ferland}, G.~J., {et~al.} 2007, \mnras,
  382, 1246

\bibitem[{{Lee} {et~al.}(1996){Lee}, {Herbst}, {Pineau des Forets}, {Roueff},
  \& {Le Bourlot}}]{Lee96}
{Lee}, H., {Herbst}, E., {Pineau des Forets}, G., {Roueff}, E., \& {Le
  Bourlot}, J. 1996, \aap, 311, 690

\bibitem[{{Lepp} \& {Dalgarno}(1996)}]{Lepp96}
{Lepp}, S. \& {Dalgarno}, A. 1996, \aap, 306, L21

\bibitem[{{Lepp} \& {Tin{\'e}}(1998)}]{Lepp98}
{Lepp}, S. \& {Tin{\'e}}, S. 1998, The Molecular Astrophysics of Stars and
  Galaxies, edited by Thomas W.~Hartquist and David A.~Williams.~Clarendon
  Press, Oxford, 1998., p.489, 4, 489

\bibitem[{{Lim} {et~al.}(2008){Lim}, {Ao}, \& {Dinh-V-Trung}}]{Lim08}
{Lim}, J., {Ao}, Y., \& {Dinh-V-Trung}. 2008, \apj, 672, 252

\bibitem[{{McNamara} {et~al.}(1996){McNamara}, {O'Connell}, \&
  {Sarazin}}]{McNa96}
{McNamara}, B.~R., {O'Connell}, R.~W., \& {Sarazin}, C.~L. 1996, \aj, 112, 91

\bibitem[{{Pope} {et~al.}(2008{\natexlab{a}}){Pope}, {Hartquist}, \&
  {Pittard}}]{Pope08a}
{Pope}, E.~C.~D., {Hartquist}, T.~W., \& {Pittard}, J.~M. 2008{\natexlab{a}},
  \mnras, 389, 1259

\bibitem[{{Pope} {et~al.}(2008{\natexlab{b}}){Pope}, {Pittard}, {Hartquist}, \&
  {Falle}}]{Pope08b}
{Pope}, E.~C.~D., {Pittard}, J.~M., {Hartquist}, T.~W., \& {Falle}, S.~A.~E.~G.
  2008{\natexlab{b}}, \mnras, 385, 1779

\bibitem[{{Rawlings} {et~al.}(2002){Rawlings}, {Hartquist}, {Williams}, \&
  {Falle}}]{Rawl02}
{Rawlings}, J.~M.~C., {Hartquist}, T.~W., {Williams}, D.~A., \& {Falle},
  S.~A.~E.~G. 2002, \aap, 391, 681

\bibitem[{{R{\"o}llig} {et~al.}(2007){R{\"o}llig}, {Abel}, {Bell}, {Bensch},
  {Black}, {Ferland}, {Jonkheid}, {Kamp}, {Kaufman}, {Le Bourlot}, {Le Petit},
  \& {etc, }}]{Roel07}
{R{\"o}llig}, M., {Abel}, N.~P., {Bell}, T., {et~al.} 2007, \aap, 467, 187

\bibitem[{{Salom{\'e}} {et~al.}(2006){Salom{\'e}}, {Combes}, {Edge},
  {Crawford}, {Erlund}, {Fabian}, {Hatch}, {Johnstone}, {Sanders}, \&
  {Wilman}}]{Salo06}
{Salom{\'e}}, P., {Combes}, F., {Edge}, A.~C., {et~al.} 2006, \aap, 454, 437

\bibitem[{{Salom{\'e}} {et~al.}(2008{\natexlab{a}}){Salom{\'e}}, {Combes},
  {Revaz}, {Edge}, {Hatch}, {Fabian}, \& {Johnstone}}]{Salo08b}
{Salom{\'e}}, P., {Combes}, F., {Revaz}, Y., {et~al.} 2008{\natexlab{a}}, \aap,
  484, 317

\bibitem[{{Salom{\'e}} {et~al.}(2008{\natexlab{b}}){Salom{\'e}}, {Revaz},
  {Combes}, {Pety}, {Downes}, {Edge}, \& {Fabian}}]{Salo08a}
{Salom{\'e}}, P., {Revaz}, Y., {Combes}, F., {et~al.} 2008{\natexlab{b}}, \aap,
  483, 793

\bibitem[{{Sanders} \& {Fabian}(2007)}]{Sand07}
{Sanders}, J.~S. \& {Fabian}, A.~C. 2007, \mnras, 381, 1381

\bibitem[{{Sanders} {et~al.}(2004){Sanders}, {Fabian}, {Allen}, \&
  {Schmidt}}]{Sand04}
{Sanders}, J.~S., {Fabian}, A.~C., {Allen}, S.~W., \& {Schmidt}, R.~W. 2004,
  \mnras, 349, 952

\bibitem[{{Sch{\"o}ier} {et~al.}(2005){Sch{\"o}ier}, {van der Tak}, {van
  Dishoeck}, \& {Black}}]{Scho05}
{Sch{\"o}ier}, F.~L., {van der Tak}, F.~F.~S., {van Dishoeck}, E.~F., \&
  {Black}, J.~H. 2005, \aap, 432, 369

\bibitem[{{van der Tak} {et~al.}(2007){van der Tak}, {Black}, {Sch{\"o}ier},
  {Jansen}, \& {van Dishoeck}}]{VanderTak07}
{van der Tak}, F.~F.~S., {Black}, J.~H., {Sch{\"o}ier}, F.~L., {Jansen}, D.~J.,
  \& {van Dishoeck}, E.~F. 2007, \aap, 468, 627

\end{thebibliography}

\section*{Acknowledgments}

EB acknowledges financial support from the Leverhulme Trust. We thank the referee for the useful comments.

\end{document}